\documentclass[letterpaper, 10 pt, conference]{ieeeconf}  

\IEEEoverridecommandlockouts                              

\usepackage{graphics} 
\usepackage{epsfig} 
\usepackage{mathptmx} 
\usepackage{times} 
\usepackage{amsmath} 
\usepackage{amssymb}  
\usepackage{multirow}
\usepackage{placeins}

\newcommand{\specialcell}[2][c]{%
  \begin{tabular}[#1]{@{}c@{}}#2\end{tabular}}

\title{\LARGE \bf Physical and behavioral comparison of haptic touchscreens quality}

\author{Corentin Bernard$^{1,2}$, Nicolas Huloux$^{2}$,  Micha\"el Wiertlewski$^{3}$
       and  Jocelyn Monnoyer$^{4}$
\thanks{This work was conducted in the framework of the Openlab PSA-AMU "Automotive Motion Lab" and the France Relance program.}
\thanks{$^{1}$Corentin Bernard is with Aix Marseille Univ, CNRS, PRISM, Marseille, France.
{\tt\small bernard@prism.cnrs.fr}}
\thanks{$^{2}$Corentin Bernard and Nicolas Huloux are with MIRA, Aflokkat, Ajaccio, France.}
\thanks{$^{3}$Micha\"el Wiertlewski is with TU Delft, Delft, The Netherlands.}
\thanks{$^{4}$Jocelyn Monnoyer is with Stellantis, Paris, France and Aix Marseille Univ, CNRS, ISM, Marseille, France. }
}

\begin{document}

\maketitle
\thispagestyle{empty}
\pagestyle{empty}

\begin{abstract}
Touchscreens equipped with friction modulation can provide rich tactile feedback to their users.
To date, there are no standard metrics to properly quantify the benefit brought by haptic feedback.
The definition of such metrics is not straightforward since friction modulation technologies can be achieved by either ultrasonic waves or with electroadhesion. In addition, the output depends strongly on the user, both because of the mechanical behavior of the fingertip and personal tactile somatosensory capabilities.
This paper proposes a method to evaluate and compare the performance of haptic tablets on an objective scale. The method first defines multiple metrics using physical measurements of friction and latency. The comparison is completed with metrics derived from information theory and based on pointing tasks performed by users.
We evaluated the comparison method with two haptic devices, one based on ultrasonic friction modulation and the other based on electroadhesion. 
This work paves the way toward the definitions of standard specifications for haptic tablets, to establish benchmarks and guidelines for improving surface haptic devices.
\end{abstract}

\section{Introduction}

Haptic feedback on touchscreens creates tactile sensations that can render human-computer interfaces tangible. With haptic feedback, the interface becomes more intuitive to use and requires less visual attention~\cite{bernard2022eyes}.

There are mainly two categories of haptic touchscreen technologies. The most common one uses low-frequency vibrations (below 800~Hz) that propagate through the plate to provide vibrotactile feedback to the user's finger~\cite{poupyrev2003tactile,pantera2020multitouch}.
This paper focuses on the second category: haptic technologies that affect the frictional forces as users slide their finger across the touchscreen. Friction modulation can be achieved by different physical principles: (i) via ultrasonic friction modulation which relies on ultrasonic levitation to reduce the friction~\cite{biet2007squeeze,wiertlewski2016partial} or (ii) via electroadhesion which relies on electrostatic attraction to increase the friction of the finger~\cite{bau2010teslatouch, shultz2018application, vardar2017effect}. More recently, it has been shown that friction can also be slightly modulated through temperature changes~\cite{choi2022surface}. 

These technologies are recent, and are confined to laboratories or startups. The generalization of haptic tablets requires that the technical specifications and performance of the haptic feedback be objectively measured and possibly standardized.
However, it remains a challenge to define performance indicators since the evaluation of the haptic feedback quality is not straightforward.
First, friction modulation technologies are highly dependent on the user's finger mechanical behaviour~\cite{pasumarty2011friction,monnoyer2018perception}. Second, the tactile sensory system and its acuity to perceive friction stimulation also play an important role~\cite{bernard2020detection}. 
Finally, other technical parameters, such as the latency or fluidity of the tablet, influence the user interaction~\cite{okamoto2009detectability}.
To properly evaluate these devices and compare them, user-in-the-loop measurements are needed.

Yet, measurements involving humans raise other issues in terms of variability and repeatability, which need to be addressed. 
Similar questions have already been considered for the comparison of force-feedback haptic devices. Although a large number of physical measures of performance can be defined~\cite{hayward1996performance,samur2012performance}, they only partially reflect the usability \cite{bowman2002survey} of the force feedback devices. Therefore, Samur proposed to assess the performances of the interfaces through the performances of users in completing a set of tasks with the device \cite{samur2012performance}.

Similarly, we propose here an evaluation method for haptic touchscreen comparison composed of two parts. First, friction measurements using users' fingers exhibit physical performance metrics. Second, the performance of users in a pointing task provides behavioral comparison metrics.
We experimentally assess the validity of the evaluation method by comparing two haptic tablets, a T-pad based on ultrasonic friction modulation built in our lab, and a Tanvas tablet based on electroadhesion. The results of the experiment are used to select the most relevant benchmark metrics.
 
\section{Description of the two evaluated tablets}

\begin{figure}[h!]
\centering
\includegraphics[]{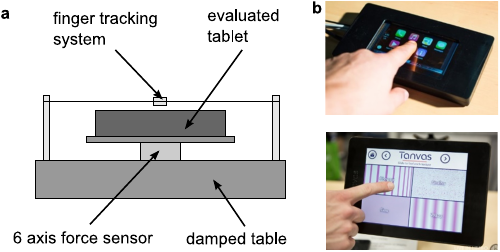}
.\caption{\textbf{a.} Schematics of the testbed for the physical comparison. \textbf{b.} Pictures of the two tablets used here to assess the comparison method: (top) the T-pad build in the lab using ultrasonic friction modulation and (bottom) the Tanvas tablet using electroadhesion (\textit{https://tanvas.co})}
\label{fig:testbed}
\end{figure}

The comparison method is evaluated on two tablets presented in Fig.~\ref{fig:testbed}.b. The first one is a T-pad based on ultrasonic friction modulation built in the lab.  A glass plate vibrates at 36~kHz to produce ultrasonic levitation and reduce friction. The finger position is measured with a laser light based sensor (zForce AIR Touch Sensor, Neonode) and the actuation is controlled by a microcontroller (Teensy 3.5, PJRC). The visual display is controlled by a computer connected to the microcontroller via serial communication.
The other device is a TanvasTouch tablet (2018) using electrovibration. It is based on an android tablet enhanced with a haptic touchscreen developed by Tanvas (Chicago, US).

\section{Physical measurements}

Firstly, basic physical measurements of friction are required to evaluate the haptic surface capability. Highest and lowest constant friction levels provide the friction range metric. It reflects the maximal possible intensity of the haptic feedback. The perception of elementary stimuli such as edges is indeed directly linked to friction change amplitude~\cite{saleem2019psychophysical,gueorguiev2017tactile,messaoud2016relation}.
This section presents the testbed and the signal analysis process to measure the physical metrics, and then applies the method to the evaluation of the two haptic tablets.

\subsection{Description of the testbed}

The testbed, presented in Fig. \ref{fig:testbed}, is composed of a damped table (Thorlabs) on which a 6-axis force sensor (Nano 43, ATI, Apex) is fixed to measure frictional forces.
On the top of the sensor, a support with two clamps ensures a rigid connection between the force sensor and the haptic tablet we want to evaluate.
Although the evaluated tablets offer their own finger tracking system, we prefer to measure the finger position externally to ensure consistent comparisons and repeatability.
A small ring attached to the participant's finger is connected to a pulley-encoder system (KIS40, Kübler) that measures unidirectional finger displacements along the length of the screen. 
The precision of this system is approximately 0.01~mm without any significant latency. 
Frictional forces and finger positions are recorded with an acquisition card (USB X Series Multifunction DAQ, National Instruments) at a 10~kHz sampling rate.

\subsection{Protocol and signal processing}

\begin{figure}[h]
\centering
\includegraphics[width=8cm]{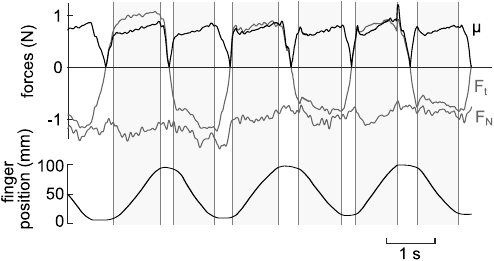}
\caption{Forces and finger position measurements from a typical trial. The finger position is used to cut the signal in order to  keep only 6 sections when the finger movement is approximately constant.}
\label{fig:signalProcessing}
\end{figure}

Ten participants took part in the physical measurements, 2 females and 8 males, from 22 to 41 years old (mean: 25.2). The study was approved by the Ethical Committee of Aix-Marseille Universit\'e. The tablet screens were cleaned with an alcoholic solution and the participants washed and dried their hands 5 minutes before starting the experiment.
All participants performed one measurement session with the T-pad and one with the Tanvas.
On each trial, they were asked to slide their right finger from left to right and right to left on the screen during 10~s, synchronising their movement with a metronome to ensure a constant velocity of about 100~mm/s.
Some trials without recording were used at the beginning to train the participants to keep their finger normal force between 0.5 and 1.5~N.

Normal force $F_N$, tangential force $F_T$ and finger position $x$ measured for one typical trial are presented in Fig. \ref{fig:signalProcessing}. The friction coefficient $\mu$ is computed as the ratio between the tangential force and the normal force $\mu=F_T/F_N$. Finger position data are used to select only the sections between two sliding direction changes and express the friction coefficient as a function of the position $x$.
For each condition of tablet and actuation, each participant performed 3 times the 10~s finger exploration with 6 finger swipes, which led to 18 measurement repetitions per participant.

The signals from the right to left finger swipes were flipped to be treated together with left to right swipes.
It appeared that all friction signals presented a constant trend with a slight increase between the start of a slide and its end. We assumed that this trend was due to mechanical crosstalk of the 6-axis force sensor. Linear regressions were performed on all trials and the mean slope ($a=0.0036~\mathrm{mm}^{-1} $) was used to correct the trend by subtracting the corrective function $\epsilon=a(x-50)$ from the friction signals.

Force signals from 4 participants showed very high variability, even without actuation, due to stick-slip effects of the finger on the glass.
Since the signals were too noisy for the analysis, their measurements were discarded from the physical comparison.

\subsection{Constant actuation and friction range}

\begin{figure}[h]
\centering
\includegraphics[]{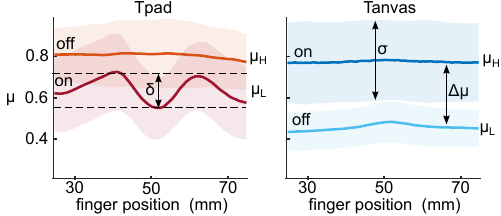}
\caption{Evaluation of the constant friction levels $\mu_H$ and $\mu_L$, the friction range $\Delta_\mu$, the inter-participant standard deviation $\sigma$ and the mean intra-trial standard deviation $\delta$. Measurements are presented for the T-pad and the Tanvas tablet for all participants. The solid lines represent the mean and the shaded zones represent the standard deviation.
}
\label{fig:constantFriction}
\end{figure}

Friction range is an important metric to assess friction changes capability of the haptic tablet. It reflects how strong the haptic effects (like edges) can be perceived~\cite{saleem2018psychophysical}.

This parameter is derived from the measurement of the highest and lowest friction levels $\mu_H$ and $\mu_L$, i.e with constant maximal actuation and without actuation. ($\mu_H$ will be obtained without actuation for the T-pad, and with constant actuation for the Tanvas). The friction range is calculated as $\Delta \mu = \mu_H - \mu_L $ as presented in Fig. 2.a. 

Constant friction measurements made on the two haptic touchscreens are presented in Fig~\ref{fig:constantFriction}.
For the T-pad with actuation, we observed curves with a sinusoidal shape. This inconstancy is due to the technology of ultrasonic friction modulation and its plate vibration nodes and antinodes \cite{marchuk2010friction}.
This highlights an important property that is the ability to provide steady stimulation to the user to render a sensation of "flatness". We therefore defined a metric assessing the variability within a finger swipe expressed by the mean friction intra-trial standard deviation $\delta$.
We also observed large differences of constant friction levels between participants. Even if these disparities are mainly due to external factors (user's finger mechanical behaviour, its angle and moisture~\cite{pasumarty2011friction}), we still propose to evaluate this aspect since some tablets could features technological solution to reduce the friction variability between users (glass processing or friction control~\cite{huloux2018overcoming} for example). Thus, we defined another metric, the inter-participant standard deviation $\sigma$ that reflects the ability to provide repeatable stimulation across users.
The measured metrics are presented in Table~\ref{tab:physical} for the two evaluated devices.
The Tanvas tablet shows a higher friction range (Two-sample T-test: $T_{214}=-8.74$, $p=7.0e^{-16}$) but the inter-participant variabilities are not statistically different (F-Test for Equality of Two Variances: $F_{107,107}=0.853$, $p=0.413$).

\subsection{End-to-end latency measurement}

\begin{figure}[h]
\centering
\includegraphics[]{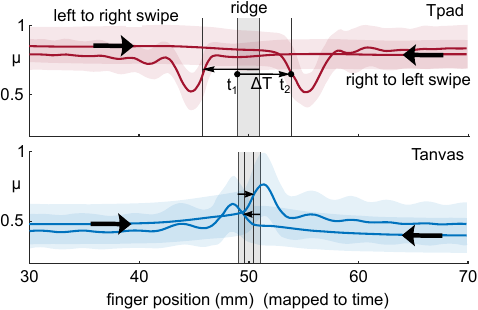}
\caption{Evaluation of the end-to-end latency $\Delta T$ trough the rendering of a haptic ridge (actuation wanted on the grey area). Friction is presented for left-to-right swipes and for right-to-left swipes to exhibit the impact of the delay. 
Friction measurements are performed on the T-pad and the Tanvas tablet for all participants. The solid lines represent the mean and the shaded zones represent the standard deviation.
}
\label{fig:latency}
\end{figure}

The end-to-end latency corresponds to the delay between a user’s action and the system final response~\cite{casiez2017characterizing}. Here we are dealing with the delay between the user's finger detection and the haptic actuation. 
This latency is due to several factors: the finger position measurement refresh rate, communication delays, running time of the microcontroller and the actuation duration (plate frequency response for the T-pad and plate charging time for the Tanvas, both impacted by the finger mechanics~\cite{meyer2014dynamics}).
A low end-to-end latency is crucial for touch based HCI to render trustful haptic feedback~\cite{okamoto2009detectability,kaaresoja2011playing}. This problem is particularly pointed out for haptic feedback that needs to be precisely located in space, like boundaries or ridges. For example, on a tablet with a 100~ms end-to-end latency, a ridge explored with a 200~mm/s left-to-right finger swipe will be felt 2~cm too far to the right, and 2~cm too far to the left when sliding in the other direction, which causes a haptic shift that strongly reduces the realism.

The end-to-end latency is here measured using this principle. A 2~mm vertical ridge is spatially programmed on the tablet, which means that the haptic actuation should activate when the finger is tracked on the ridge (a low friction ridge for the T-pad and a high friction ridge for the Tanvas).
Fig.~\ref{fig:latency} shows how the end-to-end latency $\Delta T=t_2-t_1$ is estimated by examining the instants $t_1$ when the finger crosses the ridge and $t_2$ when the actuation emerges. We defined $t_2$ as the point of inflection when the friction derivative reaches its extremum, which is the middle point between the point where the actuation starts and the point where it reaches its maximum, to take into account the actuation duration needed for the haptic feedback to be noticeable.
The results for the two tablets are reported in Table~\ref{tab:physical}. The standard deviation reports the uncertainty due to the measurement method.

\begin{table}[h]
\caption{Comparison metrics for the physical evaluation}

\begin{center}
\begin{tabular}{cl|c|c|}
\cline{3-4}
                                                &      & T-pad  & Tanvas  \\ 
                                                &      &  {\scriptsize (n=6x18)} & {\scriptsize (n=6x18)} \\ \hline 
\multicolumn{1}{|c|}{\multirow{3}{*}{Highest friction $\mu_H$}}      & Mean                &   0.771     &       0.744      \\ 
\multicolumn{1}{|l|}{}                                               & Inter-part. std $\sigma$ &      0.113     &     0.142        \\
\multicolumn{1}{|l|}{}                                                 & Intra-trial std  $\delta $ &    0.028      &     0.025        \\   \hline
                                                
\multicolumn{1}{|c|}{\multirow{3}{*}{Lowest friction  $\mu_L$}}      & Mean               &  0.620         &    0.443         \\  
\multicolumn{1}{|l|}{}                                                & Inter-part. std $\sigma$ &  0.100          &   0.079           \\ 
\multicolumn{1}{|l|}{}                                               & Intra-trial std  $\delta$  &  0.088          &    0.018         \\ \hline

\multicolumn{1}{|c|}{\multirow{2}{*}{Friction range $\Delta \mu$}} & Mean            &      0.151      &      0.301       \\ 
\multicolumn{1}{|l|}{}                                               & Inter-part. std  &     0.121      &     0.131        \\ \hline
\multicolumn{1}{|c|}{End-to-end latency $\Delta$T}                  &      Mean $\pm$ std  &   $33 \pm 3$ ms       &    $6 \pm 3$ ms      \\ \hline
\end{tabular}
\end{center}
\label{tab:physical}
\end{table}

\FloatBarrier 

\section{Behavioral measurements}

\subsection{Principle}

In the previous section, we proposed to assess the potential of haptic tablets through physical measurements.
However, since these interfaces are intended to be used by humans, the evaluation must be complemented by behavioral measurements. 
Inspired by the literature~\cite{samur2012performance}, we propose here to evaluate the performances of the haptic tablets through the performances of users in a one-dimensional pointing task: the user has to reach a target as quickly as possible.
This classical HCI task is well described under the Fitts' law framework~\cite{fitts1954information}, a predictive model of human movement that describes the trade-off between precision and rapidity in pointing at a target.
Fitts’ law predicts the average movement time as $MT=a+b\times ID$, with $a$ and $b$ constants that depend on the interface and $ID$ the index of difficulty. $ID$ is expressed for interfaces by the Shannon formulation~\cite{mackenzie1992fitts} as $ID=\mathrm{log_2}(\frac{D}{W}+1)$, with $D$ the distance to the target and $W$ the width of the target (see Fig.~\ref{fig:pointingTask}).

Many studies have demonstrated that the addition of haptic feedback using friction modulation significantly improves the performance of pointing tasks~\cite{casiez2011surfpad, levesque2011enhancing,kalantari2018determining,zhang2015quantifying}. This is mainly shown by a reduction of the movement times, reflected by a diminution of Fitts' slope $b$.

In this section, we propose to perform the same pointing task on the tablets (with a standardized protocol) with and without haptic feedback in order to evaluate the overall usability of the device and the gain of haptics.

\subsection{Protocol}

\begin{figure}[h]
\centering
\includegraphics[]{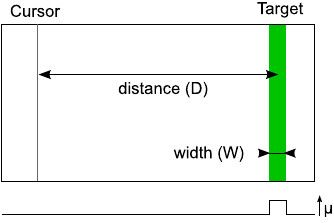}
\caption{Interface of the pointing task. }
\label{fig:pointingTask}
\end{figure}

For the behavioral comparison, 10 participants, 3 females and 7 males, from 22 to 46 years old (mean: 26.4) took part of the experiment. Half of the participants also participated in the physical measurements.
The pointing task was performed in a 100~x~60~mm window presented in Fig.~\ref{fig:pointingTask}. 
At each trial, participants were asked to select the cursor with the index of their dominant hand  and drag it into the green target as quickly as possible. The direction (right to left or left to right) was alternated at each trial.
The distance $D$ between the cursor and the target was fixed at 80~mm. There were 8 target width $W$ conditions: 1, 2, 3, 4, 5, 6, 7 and 8~mm. It led to 8 difficulty indexes $ID=\mathrm{log_2}(\frac{D}{W}+1)$: 6.3, 5.3, 4.8, 4.4, 4.1, 3.8, 3.6 and 3.4, repeated each 6 times.
For the haptic condition, the tablet was actuated to deliver a high friction in the target and a constant low friction elsewhere.
The experiment was performed with and without haptic feedback for the two tablets and the presentation order was balanced among participants. Overall, each participant performed 2 tablets x 2 actuation x 8 $ID$ x 6 repetitions = 192 trials.

\subsection{Analysis}

At each trial, the movement time ($MT$) is calculated as the time between the participant touching the cursor and releasing it. 
For each participant, the movement time is averaged over the repetitions and linear regression  $MT=a+b\times ID$ are performed to exhibit Fitts' slope $b$ used for the comparison.
The error rate is calculated by considering trials in which the cursor is released outside of the target.

\begin{figure}[h]
\centering
\includegraphics[]{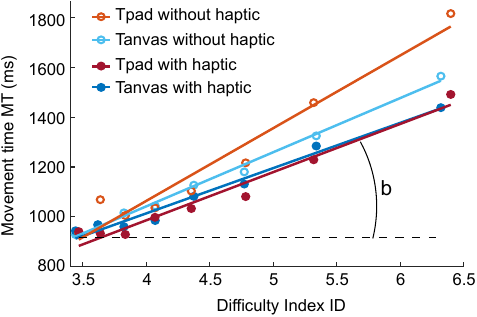}
\caption{Results of the pointing experiment. Mean movement time is plotted with respect to the difficulty index for the 4 interface conditions. Linear regressions $MT=a+b\times ID$ are calculated to exhibit Fitts' slope $b$. }
\label{fig:FittsResults}
\end{figure}

\subsection{Results}

Linear regression exhibited that the results were well in line with Fitts' law (goodness of fit $R^2 \subset [0.95 ,0.97]$). Therefore, we propose to use Fitts' slope $b$ as a first comparison metric. It reflects how much the movement time increases with the difficulty index. The Fitts' slope is calculated for each participant (n=10) and the standard deviation $\sigma$ is computed to reflect the high inter-participant variability inherent in pointing tasks. 
Instead of the Fitts' offset $a$ that is not relevant in our case, we rather propose as a second comparison metric the movement time for the most selective difficulty index ($ID=6.3$). Here, mean and standard deviation are directly calculated on the 6x10=60 trials.
Those comparison metrics are reported in Table~\ref{tab:poitingtask}.
In the present comparison, statistical analysis reported a significant effect of the interface on the movement time (for $ID=6.3$) (One-way ANOVA: $F_3=3.03$, $p=0.03$) due to the condition with the T-pad without haptic.
Performances of the Tanvas tablet with haptic (both for  movement time and error rate) were not as good as expected from the physical evaluation. We hypothesized that this effect was due to oscillations of the cursor caused by a noisier finger position measure when the electrovibration actuation was on, a problem that has since been fixed in later Tanvas prototypes.

\begin{table}[h]
\caption{Comparison metrics for the pointing task.} 

\begin{tabular}{cl|c|c|}
\cline{3-4}
\multicolumn{1}{c}{}                                   & \multicolumn{1}{c|}{} & \begin{tabular}[c]{@{}c@{}}T-pad \\ \end{tabular} & \begin{tabular}[c]{@{}c@{}}Tanvas \\ \end{tabular} \\ \hline
\multicolumn{1}{|c|}{\multirow{3}{*}{ Fitts' slope $b$}} & Without haptic        &        \specialcell{   293 ms \\  {\scriptsize($\sigma=226$, n=10)}}    &     \specialcell{  217 ms   \\  {\scriptsize ($\sigma=100$, n=10) } } \newline                               \\
\multicolumn{1}{|c|}{}                                 & With haptic           &     \specialcell{     187 ms    \\ {\scriptsize($\sigma=117$, n=10)} }                                      &              \specialcell{     180 ms  \\ {\scriptsize($\sigma=66$, n=10)}    }                              \\\hline
\multicolumn{1}{|c|}{\multirow{3}{*}{\specialcell{Movement Time\\for ID=6.3}}} & Without haptic        &    \specialcell{ 1815 ms   \\ {\scriptsize($\sigma=1045$, n=60)}     }                                         &          \specialcell{ 1564 ms\\ {\scriptsize($\sigma=562$, n=60)}}     \\  
\multicolumn{1}{|c|}{}                                 & With haptic           &     \specialcell{   1491 ms \\ {\scriptsize($\sigma=707$, n=60)}  }                                             &                  \specialcell{  1438 ms \\  {\scriptsize($\sigma=389$, n=60)}    }                               \\  \hline
\multicolumn{1}{|c|}{\multirow{2}{*}{  \specialcell{Error rate\\ (global)}}}      & Without haptic        &              10.9    \%                                    &               5.0 \%                                    \\  
\multicolumn{1}{|c|}{}                                 & With haptic           &               10.4   \%                                      &             10.4 \%                                      \\  \hline                         
\end{tabular}

\label{tab:poitingtask}

\end{table}

\section{ Discussion }
Two experiments were presented to evaluate the comparison method. The first one assessed the haptic rendering systems by measuring physical metrics relating to the friction between the screen and different user fingers. In the second one, the haptic tablets were compared on the basis of the user's performances when performing a typical human-computer interaction task.
This section discusses the results about how to improve the method and how to select a limited number of metrics to keep only the most relevant descriptors and avoid redundancy.

\subsection{Physical measurement}

The physical experiment revealed that some participants' fingers on the screen produce stick-slip effects which makes the data difficult to exploit. A first way to address this problem is to improve the testbed to reduce its resonance. The rigidity of the measurement system could be increased by adding more force sensors, one at each side for example.

Another method would be to use an artificial finger and a robotic arm for the physical measurements. It offers the advantages to be well calibrated, to provide easily replicable measurements and to strongly reduce the variability.
However, the artificial finger should present the same behaviour as human fingers both in terms of electrostatic polarizability and mechanical reaction to deformations and vibrations~\cite{friesen2015bioinspired}. Moreover, this solution does not provide any information about the variability between participants, which is an important comparison metric of consistency. Indeed, we want the tablet to be able to render the same haptic feedback to any user. In addition, a larger number of participants would be needed to reflect the variability of the population. 
The detail of the descriptors for the highest and lowest friction is of interest to investigate the sources of differences but both are included in the friction range descriptors. 
The intra-trial standard deviation is still relevant to reflect the "perceived flatness" of the surface under actuation.

The friction range metric could be improved to take into account human perception. Since the perception of friction coefficient follows a Weber law \cite{samur2009psychophysical} (with JND of friction about 20\%).
 For example, it means that the difference between $\mu_L=0.5$ and $\mu_H=0.7$ is better perceived than the difference between $\mu_L=0.8$ and $\mu_H=1$ even if the friction range is the same. Other metrics could be used like the relative friction range $r_{\mu}={\mu_H}/{\mu_L}$ or the friction contrast $FC=1-{\mu_L}/{\mu_H}$ \cite{messaoud2016relation}.

End-to-end latency is also a crucial metric that needs to be analysed in the light of human perception. Future studies on the maximal unnoticeable latency for haptic actuation could define a threshold below which the comparison is unnecessary. 

All physical metrics were measured with the same constant velocity (about 100~mm/s) and relatively steady normal force (between 0.5 and 1.5). Since frictional behaviour of the finger-glass contact is impacted by the velocity and normal force, future investigation could evaluate precisely this aspect to propose velocity and force independent metrics.

\subsection{Behavioral measurement}

The behavioral measurements take the approach of globally evaluating the tablet through the users' performance. This part is highly dependent on the participant's motor dynamics, intention and previous experience with touchscreens. Proper comparisons should therefore include a much larger number of participants than the presented preliminary experiments to reduce bias due to participant variability. It may also be worthwhile to counterbalance the effects of age with data from the literature~\cite{brogmus1991effects} or with a preliminary assessment of the participant's tactile sensitivity.

The linear regressions results demonstrated that pointing tasks with and without haptic feedback are well explained by Fitts' Law, in line with the literature \cite{casiez2011surfpad,levesque2011enhancing}.
Since the objective is to highlight differences between devices, it may not be necessary to test so many difficulty indexes. User performances could be evaluated with one target size, preferably the smallest one ($ID=6.3$) as it is the most discriminating. It would avoid redundancy between metrics and permit a much higher number of repetition to decrease variability. 

By applying the presented comparison method to a large number of haptic tablets, it would be possible to establish links between the physical and the behavioral measurement. It would be interesting to investigate how to predict the pointing performance from the physical metrics.
A better understanding of the impact of each descriptor could make it possible to apply weight to the metrics to construct an overall usability score for the haptic tablet.

\section{Summary}

The outcome of the study provided insight to improve the evaluation method. As argued in the discussion, some metrics are more relevant than others. It led us to propose an ideal comparison protocol with a limited selection of the essential metrics.
For both experiments, the panel should be constituted of 30 participants whose age follows the distribution of the adult population.
The physical comparison should be performed on a stiff testbed with force sensors on the two sides, and about twenty repetitions per measurement.
We suggest that the pointing task includes only 1 target width conditions, the most discriminatory one,  $W=1$~mm (difficulty indexes $ID=6.3$ for $D=80$~mm) and 40 repetitions.
The most relevant metrics and their descriptors are summarized in Table~\ref{tab:summary}. 
A tablet performs best when the descriptors marked with a (-) are the lowest and the descriptors marked with a (+) the highest.

\begin{table}[h]

\caption{Selection of the most relevant metrics.}

\begin{tabular}{c|cc|c|}
\cline{2-4}
                                                                                                             & \multicolumn{2}{c|}{Metrics}                                                                                  & Descriptors                                                                \\ \hline
\multicolumn{1}{|c|}{\multirow{4}{*}{\begin{tabular}[c]{@{}c@{}}Physical \\ measurement\end{tabular}}}       & \multicolumn{2}{c|}{Lowest friction}                                                                          & \multirow{2}{*}{Intra-trial SD (-)}                                        \\ \cline{2-3}
\multicolumn{1}{|c|}{}                                                                                       & \multicolumn{2}{c|}{Highest Friction}                                                                         &                                                                            \\ \cline{2-4} 
\multicolumn{1}{|c|}{}                                                                                       & \multicolumn{2}{c|}{Friction range}                                                                           & \begin{tabular}[c]{@{}c@{}}Mean (+)\\  Inter-part. SD (-)\end{tabular}     \\ \cline{2-4} 
\multicolumn{1}{|c|}{}                                                                                       & \multicolumn{2}{c|}{Latency}                                                                                  & Mean (-)                                                                   \\ \hline
\multicolumn{1}{|c|}{\multirow{4}{*}{\begin{tabular}[c]{@{}c@{}}Behavioral \\ measurement\end{tabular}}} & \multicolumn{1}{c|}{\multirow{2}{*}{\begin{tabular}[c]{@{}c@{}}$MT$\\  ($ID=6.3$)\end{tabular}}} & Without haptic & \multirow{2}{*}{\begin{tabular}[c]{@{}c@{}}Mean (-)\\ SD (-)\end{tabular}} \\ \cline{3-3}
\multicolumn{1}{|c|}{}                                                                                       & \multicolumn{1}{c|}{}                                                                        & With haptic    &                                                                            \\ \cline{2-4} 
\multicolumn{1}{|c|}{}                                                                                       & \multicolumn{1}{c|}{\multirow{2}{*}{Error rate}}                                             & Without haptic & \multirow{2}{*}{Mean (-)}                                                  \\ \cline{3-3}
\multicolumn{1}{|c|}{}                                                                                       & \multicolumn{1}{c|}{}                                                                        & With haptic    &                                                                            \\ \hline
\end{tabular}

\label{tab:summary}

\end{table}

We propose to report for each metric the mean, standard deviation and number of samples to enable anyone to easily  perform statistical comparison of their own data, with Two-Sample T-Test for example.



\section{Conclusion}

This paper is a first attempt to define objective metrics to compare different haptic tablets, even if they are based on different technologies.
The proposed method was evaluated with two different haptic devices to demonstrate its validity and to select the most relevant metrics.

Future work will investigate how the method could be extended to compare with haptic interfaces that are not based on friction modulation, such as vibrotactile tablets.
This paper lays the foundation for defining generic standards for haptic touchscreens.
It would allow consumers to objectively compare between devices and to request certain specifications. It would also enable manufacturers to analyze their devices and identify ways of improvement.

\addtolength{\textheight}{-12cm}   



\bibliographystyle{IEEEtran}
\bibliography{biblio}

\end{document}